# COMPACT USB-BASED INSTRUMENTS FOR EDUCATION AND REMOTE LABORATORY PROJECTS

**Robert Mingesz, Zoltan Gingl, Janos Mellar and Gergely Vadai,** *Department of Technical Informatics, University of Szeged*

## Abstract
Interactivity, experimenting and performing laboratory practicals contribute much to the efficiency of natural science and engineering education. Computers represent a natural element for today's students, thus the use of computers in experiments may engender considerable enthusiasm amongst them. Here we introduce two open-source, compact USB-based data acquisition devices, which we have developed for the purposes of education. These instruments offer connectivity to various sensors, like thermistors, photodiodes, sensors of acceleration, pressure, magnetic field and many other physical quantities, and provide both digital and analogue outputs.

## 1. Introduction

Efficient natural science education requires continuous development of teaching methods. In the primary and secondary school it is essential to introduce the principles of natural phenomena, to teach how to use the mathematical tools, physical and other laws to solve problems, to try to keep the pupils listening to the lecture and show also how this knowledge can be connected to everyday phenomena, modern technology, smart electronic devices. Universities are facing the problem of balancing theoretical and experimental education and methods, since the employers, companies need creative people who are not only well-educated in theoretical fields but can also use their knowledge to solve practical, sometimes unexpected problems. In addition, modern technology is in a continuous rapid development, huge information is accessible within seconds, therefore the traditional teaching methods are rarely effective, development of education is really challenging (High Level Group on Science Education 2007).

On the other hand, modern technology can help a lot to make teaching more efficient and exciting. There are many different ways of applying modern informatics, electronics and multimedia tools to help students to understand physical and other natural science phenomena including visualization, interactive learning and development of simulation software. However today's widely available advanced electronics, sensors, wireless technology, computer controlled data acquisition systems and easy-to-use software support even real-time experimenting (Comlab; Butlin 2001; Hunt 2001; Murovec and Kocijancic 2003; iSES; Vernier). Sensors translate the different physical signals into electronic signals that can be digitized and processed by computers; this way the computer can be turned into an instrument. Since most of the processing and data presentation is done by software – therefore these instruments are often called virtual instruments –, a single hardware can be used to realize incredibly large number of instruments that perform real measurements (Kántor and Gingl 2002). This kind of experimenting is very useful: students can see the processes in real time; much more can be visualized than with traditional methods; the experiment and measurement can be more transparent, more details can be shown; low price and wide availability allows the use of many instances simultaneously, the students can do some experiments in pairs or alone; the students can even make their own instruments and use their creativity to solve problems; open-source developments and software can serve as reference educational material and a basis of further developments, additions (Ganci and Ganci 2009; Gintautas and Hübler 2009; Gingl et al 2011). Another intensively developing way of experimental education is the field of remotely controlled laboratories (RCLs), where the experiment is real, but the user controls and monitors it via the internet (Eckert et al 2009; Jara et al 2009 ; Jara et al 2011; Schauer et al 2008; Gröber et al 2008). This way it needs small room, it is reliable, available at any time, can be viewed from anywhere and it is more natural than simulation.

We have developed compact, cheap USB data acquisition devices to support software based instrumentation both for direct and remotely controlled experimentation in various teaching levels.
The tested hardware design and software source code are made freely available on a dedicated web page to allow reproduction and provide reference design. RCLs are typically built and maintained only at a few universities; however these cheap and open-source devices can help to extend the RCL technique; teaches the principle, the solutions, electronic and software design and make the method much more transparent in the same time.

It is important to note, that rapidly changing computer hardware and operating system may cause problems in operating RCLs for extended time, since the factory-made data acquisition and control hardware and software may become incompatible. Simple enough open-source development has significant advantage, since all the hardware and software can be kept up-to-date easily.

## 2. USB data acquisition systems

In this chapter we introduce two simple data acquisition and control devices. Both can be connected to and powered from the USB port of the host computer thus reliable operation is provided for all popular computer manufacturers and operating systems. The units are based on mixed signal microcontrollers that incorporate analogue-to-digital converters (ADC) and other analogue components. The fast, popular 8051 architecture microcontrollers have very rich set of peripherals and are supported by free C compiler and integrated development environment with a low cost in-system programmer and debugger (www.silabs.com).

### *2.1. Compact 12-bit data acquisition system: EDAQ530A*

Our most compact and lowest cost device is a simplified version of the EDAQ530 unit (Kopasz et al 2011). The block diagram and the photo of the printed circuit board are shown on Figure 1. Only three integrated circuits are used: a C8051F530 microcontroller with a built-in 12-bit ADC, voltage reference and fast 8051 core; a popular and reliable FT232RL USB-UART interface; an LTC6484 quad operational amplifier. The 12-bit ADC has a maximum sample rate of 200k samples per second, the internal multiplexer can be used to select one of the three analogue inputs. Each terminal of the 8-pin connector can be used as digital input or digital output or can be selected as analogue input of the ADC. The flexible microcontroller provides popular serial interfaces to be routed to this connector as well (SPI, SMBUS, I2C, LIN) while 16-bit timers, counters, PWM generators and an analogue comparator are also available (see the C8051F530 datasheet at www.silabs.com).

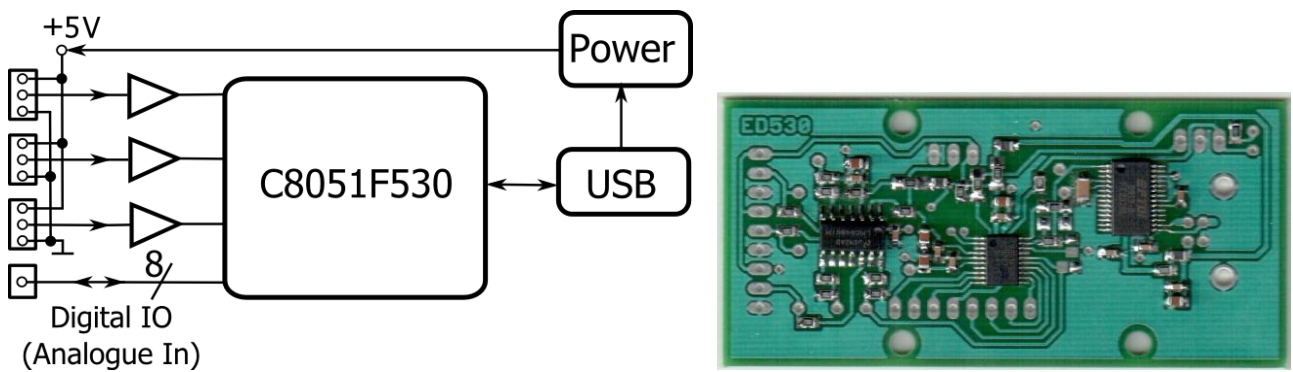

*Figure 1: Block diagram and photo of the EDAQ530A data acquisition system. Only three integrated circuits are needed, therefore the cost is kept very low and the manual assembly is rather simple.*

Three 3-pin connectors allow the connection of various sensors. The pins provide ground connection, 0V…5V voltage input and 5V power. Most sensors can be connected easily. Voltage output sensors can be just connected without any additional components if their output range matches. Too large voltages need attenuation while too small signals must be amplified. Current output sensors must be shunted by a resistor to produce voltage, and resistive sensors are typically used in an external voltage divider configuration to produce measurable voltage. Figure 2 summarizes the above mentioned connections and also lists some popular sensors.

Note that the 5V output can also be used to power external circuits, for example an amplifier. This makes possible to use very small signal sensors like thermocouples; Wheatstone-bridge sensors like pressure sensors followed by instrumentation amplifiers; very high impedance sensors like pH electrodes.

The 8-pin port allows connection of digital output sensors, for example I2C or SPI temperature, acceleration, magnetic field or pressure sensors. The internal timing unit can measure pulse width or frequency of digital signals to further extend the possibilities. Operating as digital output these signals can be used to control switches, stepper motors, to generate PWM or variable frequency periodic signals.

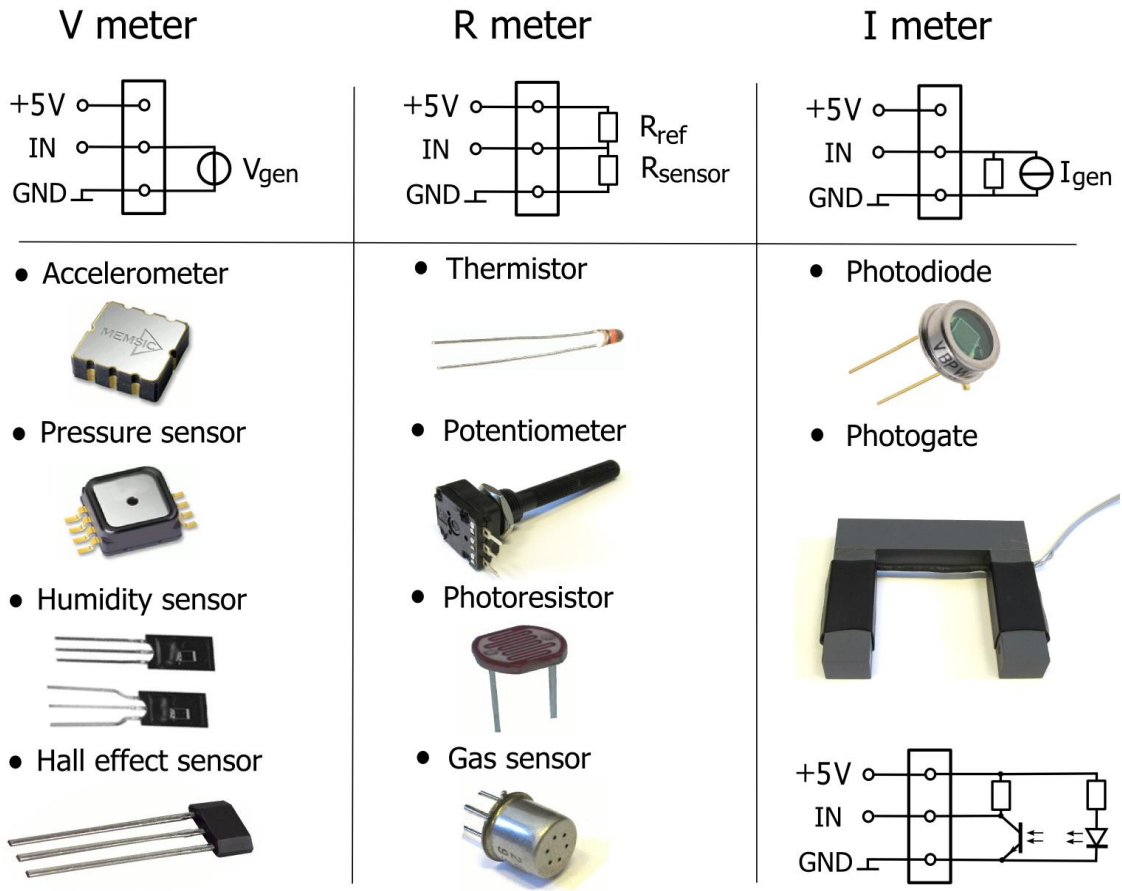

*Figure 2: Voltage and current output sensors, resistive sensors can be easily connected to the inputs if their signals meet the input range and resolution. Some tested sensors are listed below the connections diagram.*

The small size and very few components ensure easy assembly. The total cost of the components is below 50€.

### 2.2. High-performance 16-bit data acquisition system: EDAQ16

Sometimes higher resolution, bipolar signal range and more flexibility are needed therefore we have developed another USB data acquisition device that is based on a similar principle. The block diagram of the system is depicted on Figure 3.
The C8051F060 microcontroller has two built-in simultaneously sampling ADCs have 16-bit resolution and a maximum sample rate of 1M samples per second. Four inputs can be measured in the range of -5V...5V, and additional software programmable amplifiers provide gain from 1 to 128 in binary steps. The two 12.bit digital-to-analogue converters (DACs) have the same output voltage range of -5V...5V. Four general purpose digital input/output lines are available, can be programmed as simple digital I/O or as SPI, I2C serial port, PWM signal generator, event counter and more. The unit features a 512kbyte memory to store the rapidly acquired data temporarily; this ensures that not data will be lost even if the host computer can't read the data in time. The system is galvanically isolated from the computer and provides isolated -5V and 5V power as well.
Since this unit has analogue voltage inputs most of the sensors can be connected as it was shown for the EDAQ530A device. However the two independent ADCs with programmable gain amplifiers allow direct connection of thermocouples and Wheatstone bridge sensors, no additional components are needed. The very high input impedance and picoampere bias current makes it possible to directly connect pH electrodes and other high-impedance sensors as well.

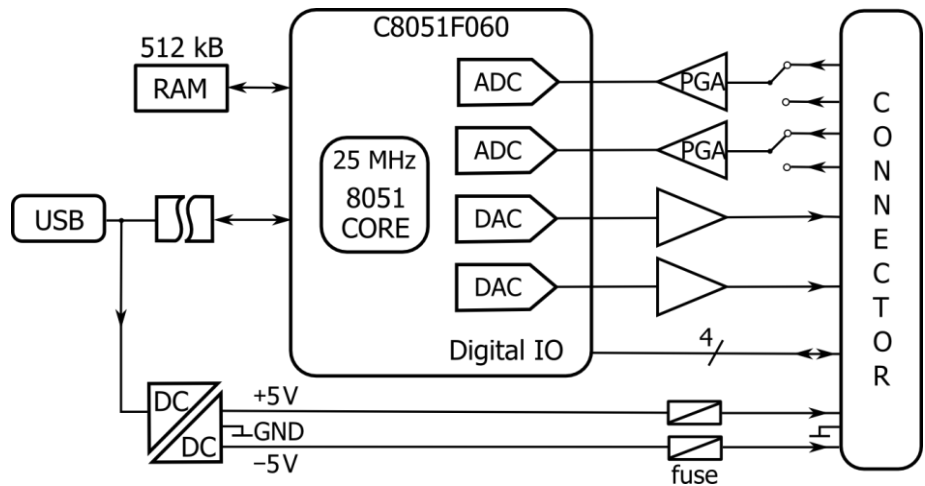

*Figure 3: Block diagram of the EDAQ16 high-performance data acquisition unit. It features two independent 1M samples/s 16-bit ADCs, 12-bit DACs, bipolar voltage range and galvanic isolation from the host.*

Due to the presence of dual analogue outputs, bipolar range, simultaneous sampling capability and high sample rate the device can be easily used to measure DC and AC transfer functions, can record transients, can be used as a two-channel oscilloscope and as an arbitrary waveform generator.

The estimated cost of the components is below 100€, and the assembly is more challenging compared to the case of the EDAQ530A, but costly manufacturing services are still avoided.

### 3. Embedded and host computer software

The microcontrollers in the devices run rather simple embedded software, since most of the processing is done by the host computer. This software can be updated by reloading the flash memory of the microcontroller using the programming adapter, but it is also possible to implement a downloader algorithm to allow reprogramming simply via the USB port.

The devices accept simple commands sent by the host computer and set the measurement and control parameters accordingly. After setting up the measurement either individual or stream of samples can be received. The microcontroller software has been written in C, and available for download (www.inf.u-szeged.hu/noise/edudev).

The host computer software can send commands and receive data via the USB port using the virtual COM port feature of the operating system or by directly accessing the driver. (www.ftdichip.com). This allows simple communication in any programming environment including C++, Java, C#, LabVIEW, Matlab/Simulink and many more on Windows, Linux, MacIntosh computers. We provide open-source C# and LabVIEW examples on a dedicated web page (www.inf.u-szeged.hu/noise/edudev)

### 4. Demonstration experiments

The developed data acquisition units have been tested and used in many different educational and scientific applications; here we show one example experiment for both units.

The first experiment demonstrates the use of the EDAQ530A unit. A Hall-effect magnetic field sensor (SS49E) was used to track the oscillation of a mass on a spring. A permanent magnet was attached to the iron mass and the sensor was placed under it. This method has several advantages compared to the use of photogates, because the instantaneous amplitude is monitored with high sensitivity. The measured waveform is shown on the left of Figure 4.

The second experiment demonstrates the use of the EDAQ16 data acquisition unit in a DC characteristics measurement. The analogue output drove a ZPY 4V7 Zener-diode via a resistor that was also used to measure the current flowing through the diode. The I-V plot can be seen on the right hand side of Figure 4.

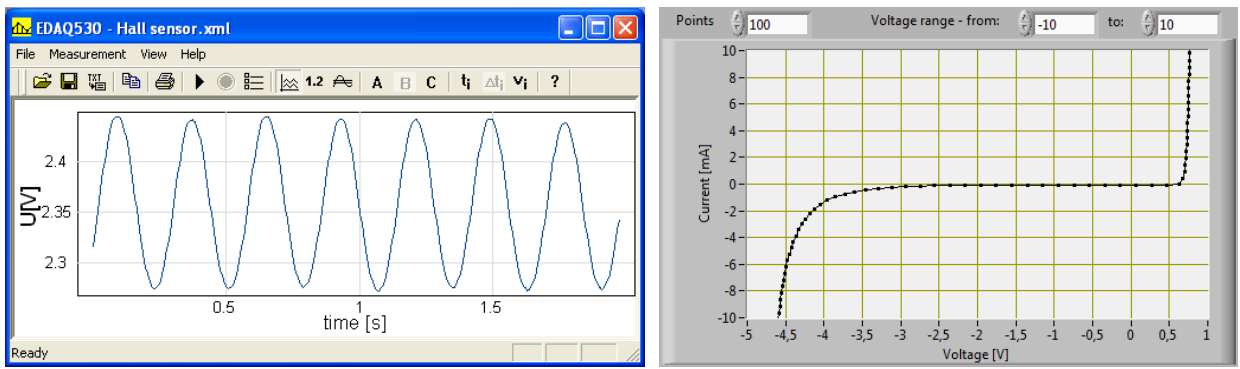

*Figure 4: Left panel: oscillation of a mass on a spring measured by a magnetic field sensor and EDAQ530A. Right panel: I-V characteristics of a Zener-diode measured by the EDAQ16 data acquisition system.*

## 5. Conclusion

We have developed two simple USB data acquisition units to support experiments at various levels of natural science education. One of the units is optimized for very low cost, easy reproduction while the other has high accuracy and more features. One of aims was to provide simple and reliable interfaces between sensors, actuators and personal computers to support reliable remotely controlled labs for long time operation. A wide range of sensors and actuators can be connected to the units with minimal effort; the units support analogue and digital input and output, timers, counters and various serial communication protocols. It is important to note that the devices can be controlled in almost any programming environment in Windows, Linux and MacIntosh computers. The developments are fully open-source; all hardware documentation and software source code are shared on the web. For more information please visit www.inf.u-szeged.hu/noise/edudev.